\documentstyle[epsf]{article}

\def\abstract#1{\vskip 7mm 
        \begin{center}{\large Abstract}\par \smallskip
                \begin{minipage}[c]{12cm}
                        \small #1
                \end{minipage}
        \end{center}
}
\def\title#1{\begin{center}{\Large\bf #1}\end{center}}
\def\author#1{\vskip 5mm \begin{center}{#1}\end{center}}
\def\address#1{\begin{center}{\it #1}\end{center}}
\makeatletter
\@ifundefined{lesssim}{}{}
\@ifundefined{gtrsim}{\def\gtrsim{\mathrel{\mathpalette\vereq>}}}{}
\def\vereq#1#2{\lower3pt\vbox{\baselineskip1.5pt \lineskip1.5pt
\ialign{$\m@th#1\hfill##\hfil$\crcr#2\crcr\sim\crcr}}}
\makeatother

\begin{document}

\title{%
  Gravitational Memory ?
  \smallskip \\
  {\large --- a Perturbative Approach  ---}
}
\author{%
  T. Harada\footnote{E-mail:harada@gravity.phys.waseda.ac.jp},
}
\address{%
  Department of Physics, Waseda University, \\
  Shinjuku, Tokyo 169-8555, Japan
}
\author{%
  B.J. Carr\footnote{E-mail:B.J.Carr@qmw.ac.uk}, and
  C.A. Goymer\footnote{E-mail:C.A.Goymer@qmw.ac.uk}
}
\address{
Astronomy Unit,
Queen Mary and Westfield College,\\University of London,
         London E1 4NS, England
}
\abstract{
It has been pointed out that the value of 
the gravitational ``constant''
in the early universe may be different 
from that at present.
In that case, it was conjectured that 
primordial black holes may ``remember'' the value of 
the gravitational constant in the early universe.
The present analysis shows that this is not the case, at least
in certain contexts.
}

\section{Introduction}
In the early universe, black holes may have been formed 
due to inhomogeneous initial conditions,
phase transitions, or other mechanisms.
They are called primordial black holes (PBHs).
The mass of the PBHs
is of order the mass contained within the
Hubble horizon at the formation epoch.
Hawking~\cite{hawking1975} 
discovered that black holes emit radiation
due to the quantum effects of the curved spacetime.
PBHs therefore lose mass and those lighter
than $\sim 10^{15}\mbox{g}$ may contribute to
the cosmic gamma-ray background
and spoil the success of 
the big bang nucleosysthesis scenario.
PBHs heavier than that
may dominate the density of the universe
at present.
Therefore the fraction $\beta$
of the universe which may have gone into PBHs
can be constrained from various cosmological observations
(Carr~\cite{carr1975}).

It has been pointed out that gravity in the early universe
may have been dilatonic, since this arises naturally
as a low-energy limit of string theory.
In such theories, the gravitational ``constant''
is given by a function of a scalar field which 
couples non-minimally with gravity.
In particular, the gravitational constant 
in the early universe
may have been different from that at the present epoch.
The Brans-Dicke theory is the simplest such theory
of gravity.
Barrow~\cite{barrow1992} 
considered PBHs in Brans-Dicke theory and
proposed two scenarios. In scenario A,
the gravitational constant at the black hole event horizon
is always the same as that at the cosmological
particle horizon, i.e., $G_{EH}(t)=G_{PH}(t)$.
In scenario B, the gravitational constant at the 
black hole event horizon is constant with time and therefore 
always the same as that at the formation epoch, i.e.,
$G_{EH}(t)=G_{EH}(t_{f})=G_{PH}(t_{f})$.  
Scenario B corresponds to what is called the {\em gravitational memory},
while scenario (A) entails no gravitational memory.

Barrow and Carr~\cite{bc1996} obtained observational constraints 
on $\beta$ in both scenarios.
They assumed that the energy emission due to the Hawking 
radiation is determined by the temperature $T_{H}=(8\pi G_{EH}(t)M)^{-1}$,
so that the mass loss rate is $\dot{M}\simeq -(G_{EH}(t)M)^{-2}$.
The results for the two scenarios can be significantly different.
Subsequently, Carr and Goymer~\cite{cg2000} argued that
the true situation may be intermediate between Scenario A and 
Scenario B and this allows other possibilities. Therefore, in order to 
deduce the history of the early universe from
the present observations,
it is very important to examine which scenario is realized.

Recently, Jacobson~\cite{jacobson1999} has argued that there is 
no gravitational memory for a Schwarzschild black hole
with a time-varying boundary condition
which mimics the cosmological evolution of a 
scalar field. However, this argument applies only if the black
hole is much smaller than the particle horizon, so we must
consider more general situations.
The purpose of this article is to solve the problem
self-consistently for a 
Friedmann-Robertson-Walker (FRW) background.

\section{Basic Equations}

The field equations in Brans-Dicke theory are given by
\begin{eqnarray}
& &R_{\mu\nu}-\frac{1}{2}g_{\mu\nu}R=\frac{8\pi}{\phi}
T_{\mu\nu}+\frac{\omega}{\phi^{2}}\left(\phi_{,\mu}
\phi_{,\nu}-\frac{1}{2}g_{\mu\nu}\phi_{,\alpha}
\phi^{,\alpha}\right)
+\frac{1}{\phi}\left(\nabla_{\mu}\nabla_{\nu}\phi
-g_{\mu\nu}\Box \phi\right), 
\label{eq:FE1}\\
& &\nabla_{\mu}T^{\mu\nu}=0, 
\label{eq:FE2}\\
& &\Box\phi=\frac{8\pi}{3+2\omega}T^{\mu}_{\quad\mu},
\label{eq:FE3}
\end{eqnarray}
where the constant $\omega$ is the Brans-Dicke parameter
and $\phi$ is the Brans-Dicke scalar field, which is related to
the gravitational constant by $G\sim \phi^{-1}$.
In the limit $\omega \to \infty$, the theory goes to
general relativity with a minimally coupled scalar
field. 
The constraint
$\omega\gtrsim 3300$ is required by recent observations of
light deflection~\cite{will1999}. However, if we consider more general scalar-tensor
theories, $\omega$ also becomes a function of $\phi$.
In this case, $\omega$ may have been small in the early universe, even though it is
large today.

Here we calculate the time evolution of the gravitational constant
by using the following post-general-relativistic expansion.
First we assume that the scalar field is constant,
i.e., $\phi=\phi_{0}$.
Then we find the general relativistic solution 
$(g_{GR}, T_{GR})$ which
satisfies Eqs. (\ref{eq:FE1}) and (\ref{eq:FE2})
with $G=\phi_{0}^{-1}$. Next we put $(g,T)=(g_{GR},T_{GR})$
in Eq.~(\ref{eq:FE3}) and assume that 
$\delta\phi=\phi-\phi_{0}$
is of order $\omega^{-1}$ for $\omega\gg 1$.
We can then obtain $\phi$ to this order,
denoting it as $\phi^{(1)}$, 
by solving the following 
equation:
\begin{equation}
\Box_{GR}\phi^{(1)}=\frac{8\pi}{3+2\omega}{T_{GR}}^{\mu}_{~~\mu},
\label{eq:BE}
\end{equation}
where $\Box_{GR}$ denotes the d'Alembertian in the 
geometry given by $g_{GR}$.
Substituting the solution $\phi^{(1)}$ of Eq.~(\ref{eq:BE})
into the right-hand side of Eq.~(\ref{eq:FE1}), we can then 
determine $g^{(1)}$ and $T^{(1)}$ by solving 
Eqs. (\ref{eq:FE1}) and (\ref{eq:FE2}).
Then, $\phi^{(2)}$, which is the solution
up to $O(\omega^{-2})$, 
is determined 
by solving Eq.~(\ref{eq:FE3}) 
for the background $(g^{(1)}, T^{(1)})$.
Repeating this process, we can construct the Brans-Dicke 
solution from the general relativity solution
order by order.

Here we truncate the expansion for the scalar field 
at $O(\omega^{-1})$.
This implies that we only have to solve Eq.~(\ref{eq:BE}) and
we can neglect the effect of the scalar field on the background
curvature.
This approximation was shown to be very good for 
the generation and propagation of gravitational waves from 
the collapse to a black hole in asymptotically flat spacetime for
Brans-Dicke theory with $\omega\gtrsim 4$~\cite{hcnn1997}.

\section{Lemaitre-Tolman-Bondi solution}
We adopt the Lemaitre-Tolman-Bondi model
as the background solution $(g_{GR}, T_{GR})$
in which the scalar field evolves.
This model is an exact solution of general relativity 
which describes
a spherically symmetric inhomogeneous
universe with dust.
The line element in the synchronous comoving 
coordinates is given by
\begin{equation}
ds^{2}=-dt^{2}+A^{2}(t,r)dr^{2}+R^{2}(t,r)d\Omega^{2},
\end{equation}
where 
\begin{equation}
d\Omega^{2}=d\theta^{2}+\sin^{2}\theta d\phi^{2}.
\end{equation}
The function $R(t,r)$ is given by
\begin{eqnarray}
& &R=\left(\frac{9F(r)}{4}\right)^{1/3}
(t-t_s(r))^{2/3}\quad \mbox{for}\quad f=0, \\
& &\left\{
\begin{array}{l}
R=\displaystyle{\frac{F(r)}{2f(r)}}(\cosh\eta -1) \\
t-t_s(r)=\displaystyle{\frac{F(r)}{2f^{3/2}(r)}}(\sinh\eta-1)
\end{array}\right.
\quad\mbox{for}\quad f>0, \\
& &\left\{
\begin{array}{l}
R=\displaystyle{\frac{F(r)}{2(-f)(r)}}(1-\cos\eta) \\
t-t_s(r)=\displaystyle{\frac{F(r)}{2(-f)^{3/2}(r)}}
(\eta - \sin\eta)
\end{array}\right.
\quad\mbox{for}\quad f<0, 
\end{eqnarray}
where the prime and dot denote the derivatives
with respect to $r$ and $t$, respectively.
The function $A(t,r)$ is given by
\begin{equation}
A^{2}(t,r)=\frac{(R^{\prime})^{2}}{1+f(r)}.
\end{equation}
The functions $t_{s}(r)$, $f(r)>-1$ and $F(r)\ge 0$ are arbitrary
and one of them corresponds to the gauge freedom.
The first two relate to the big bang time and the total energy per 
unit mass for the shell at radius $r$. The third relates to the mass
within radius $r$, the density of the dust being given by
\begin{equation}
\rho=\frac{F^{\prime}}{8\pi R^{2}R^{\prime}}.
\end{equation}

\section{Method}
We are interested in the behavior of the scalar field
long after the black hole has formed.
In this case, the characteristic method is 
suitable.
This method was first applied to the Lemaitre-Tolman-Bondi
background by Iguchi, Nakao and Harada~\cite{inh1998}.
If we introduce the retarded time $u$,
the line element becomes
\begin{equation}
ds^{2}=-\alpha(u,\bar{r}) du^{2}-2\alpha(u,\bar{\alpha}) 
A(u,\bar{r}) du d\bar{r}
+R^{2}(u,\bar{r})d\Omega^{2},
\end{equation}
where $\alpha$ must satisfy 
\begin{equation}
\frac{\partial}{\partial\bar{r}}\alpha=\dot{A}\alpha.
\end{equation}
In terms of the derivatives, $d/du$ and $\partial/\partial\bar{r}$,
the d'Alembertian is given by 
\begin{eqnarray}
\Box \phi&=&-\frac{2}{\alpha A R}\frac{d\varphi}{du}
-\frac{A^{\prime}}{A^{3}R}\varphi
+\frac{1}{AR}\left[(A\dot{R})^{\cdot}
-\left(\frac{R^{\prime}}{A}\right)^{\prime}
\right]\phi, \\
\varphi&=&\frac{\partial}{\partial\bar{r}}(R\phi).
\end{eqnarray}
We can then integrate Eq.~(\ref{eq:BE})
along the characteristic curves, i.e.,
null geodesics.
With this choice of coordinate system,
we can calculate the behavior of the scalar field long 
after black hole formation
without having to impose boundary conditions
on the event horizon.

\section{Models}
As we have seen,
the Lemaitre-Tolman-Bondi solution has
three arbitrary functions. 
We want to obtain models which describe 
a PBH in a flat FRW universe by fixing
these functions appropriately.
Here we adopt the following assumptions:
the big bang occurs at the same time everywhere;
the model is asymptotically flat FRW, which requires 
the overdense region to be compensated by a surrounding underdense region;
the model is free from naked singularities.
In order to satisfy these assumptions,
we first choose
\begin{equation}
t_{s}(r)=0,
\end{equation}
and 
\begin{equation}
f(r)=\left\{
\begin{array}{l}
-\left(\displaystyle{\frac{r}{r_{c}}}\right)^{2} 
\quad \mbox{for}\quad r<r_{w} \\
-\left(\displaystyle{\frac{r}{r_{c}}}\right)^{2}
\exp\left[-\left(\displaystyle{\frac{r-r_{w}}{r_{w}}}
\right)^{4}\right] \quad \mbox{for}\quad  r\ge r_{w}
\end{array}\right. ,
\end{equation}
where $r_{c}$ and $r_{w}$ are the curvature scale 
and the size of the overdense region.
Using the arbitrary function $F(r)$, we then set the radial 
coordinate $r$ to be
\begin{equation}
r=R(t_{0},r)
\end{equation}
where $t=t_{0}$ is the initial time.

Before integrating Eq.~(\ref{eq:BE}),
we have to fix the initial data for $\phi$, so we choose
the cosmological homogeneous
solution given by
\begin{equation}
\phi=\phi_{0}\left(1+\frac{1}{3+2\omega}\frac{4}{3}
\ln \frac{t}{t_{0}}\right).
\end{equation}
We set the initial null hypersurface as the null cone
whose vertex is at $(t,r)=(t_{0},0)$
and regard the cosmological solution as the initial 
data on this
hypersurface.
Although the value of the scalar field at the 
cosmological particle horizon must be given
by this solution, the value in the perturbed region 
may be different from this.
We have therefore examined the effects of other initial data
which are different from the cosmological solution
in the perturbed region. However, the results
are not changed much. 

\section{Results}
We adopt units in which the asymptotic value of the 
Hubble parameter at $t=t_{0}$ is unity.
We set the parameters which specify the initial data
as $r_{c}=2$ and $r_{w}=1$, while the Brans-Dicke parameter 
is chosen as $\omega=5$. 
It is noted that $r_{c}$ must satisfy $r_{c}\gtrsim r_{w}$,
else the overdense region is isolated from the
rest of the universe, as Carr and Hawking~\cite{ch1974} pointed out.

\begin{figure}[htbp]
	\centerline{\epsfxsize 10cm \epsfysize 10cm \epsfbox{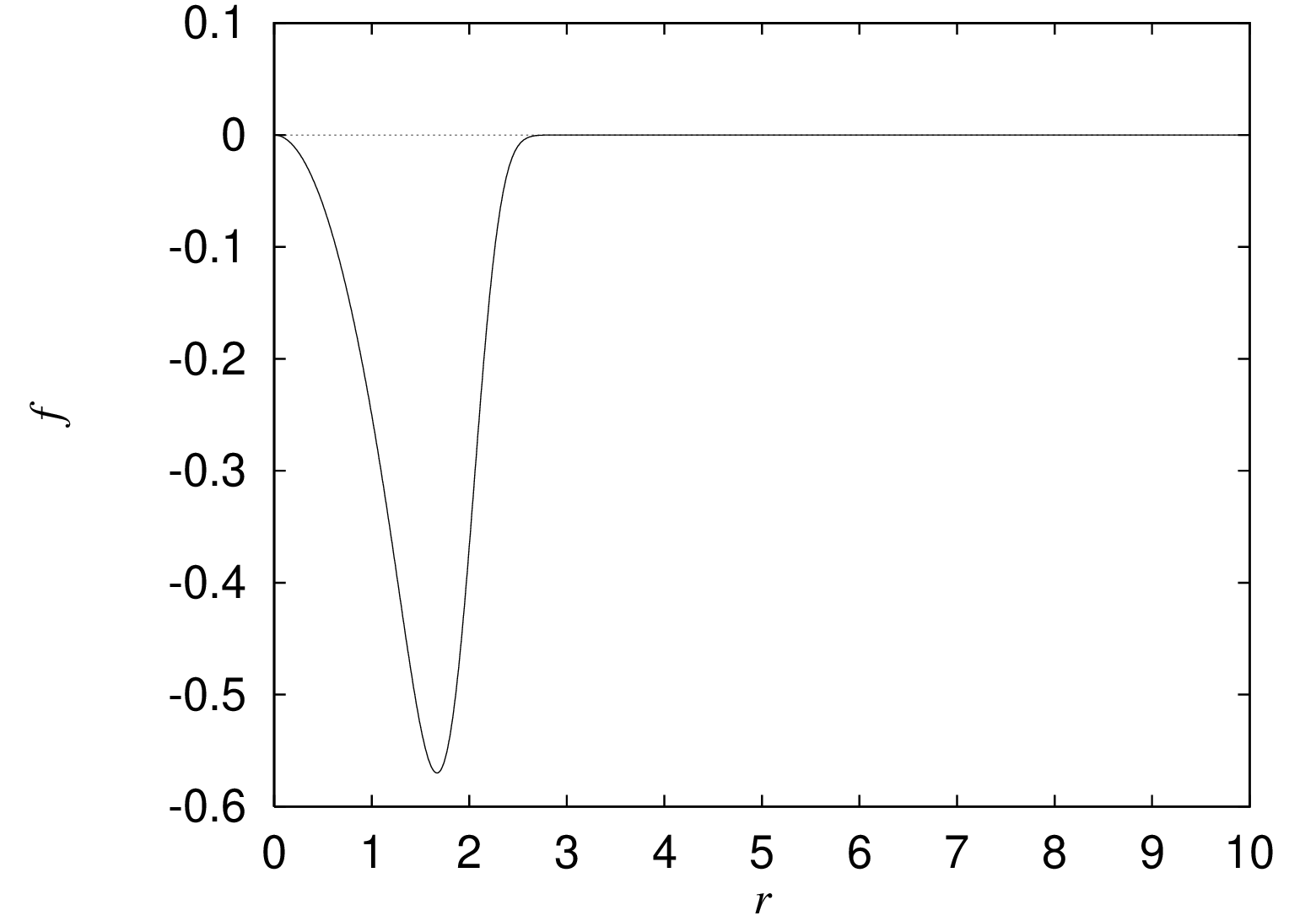}}
	(a)
	\centerline{\epsfxsize 10cm \epsfysize 10cm \epsfbox{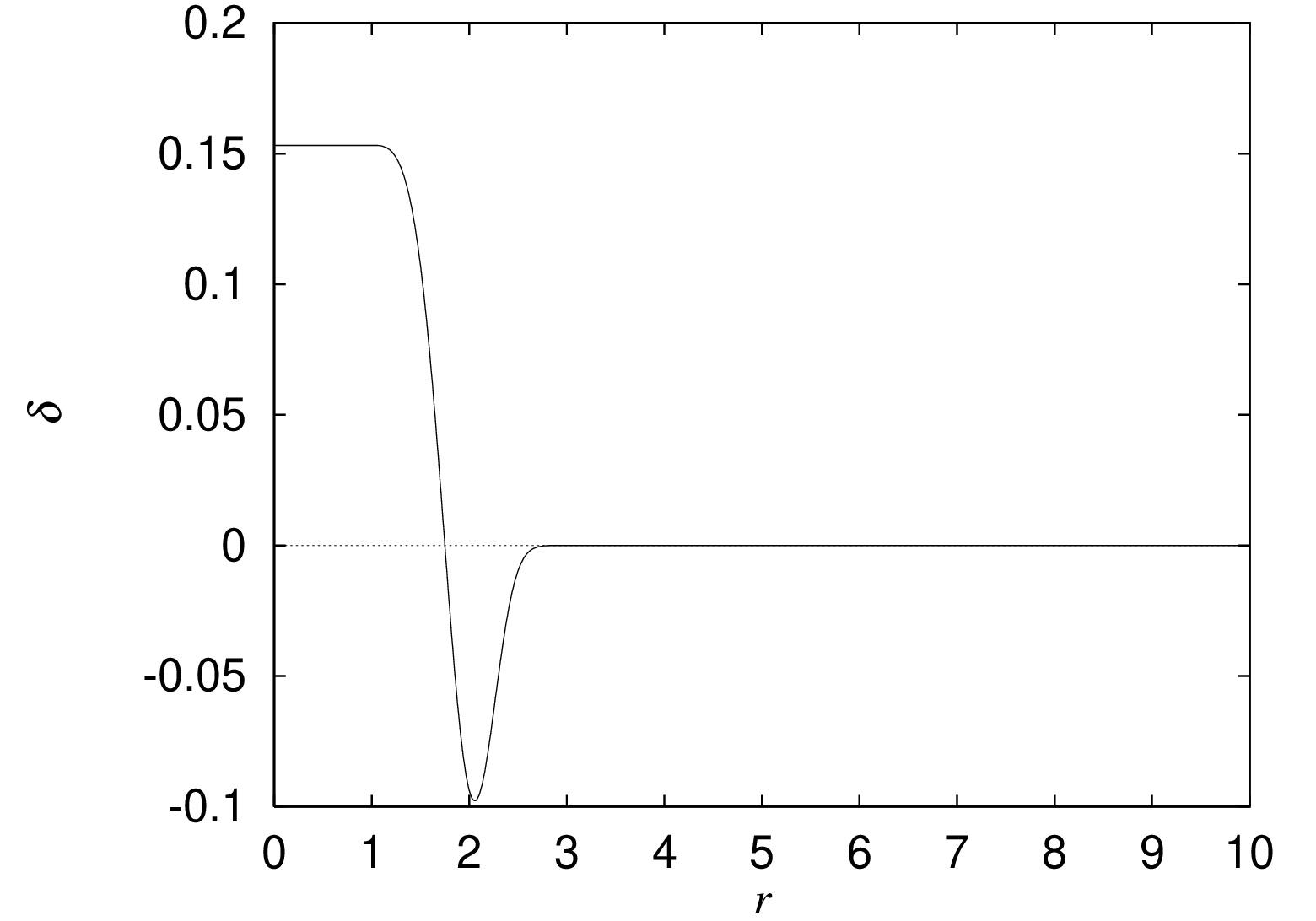}}
	(b)
	\caption{(a) Specific energy $f(r)$ and (b) initial density
	perturbation $\delta$ at $t=t_{0}$.}
	\label{fg:initial}
\end{figure}

\begin{figure}[htbp]
	\centerline{\epsfxsize 10cm \epsfysize 10cm \epsfbox{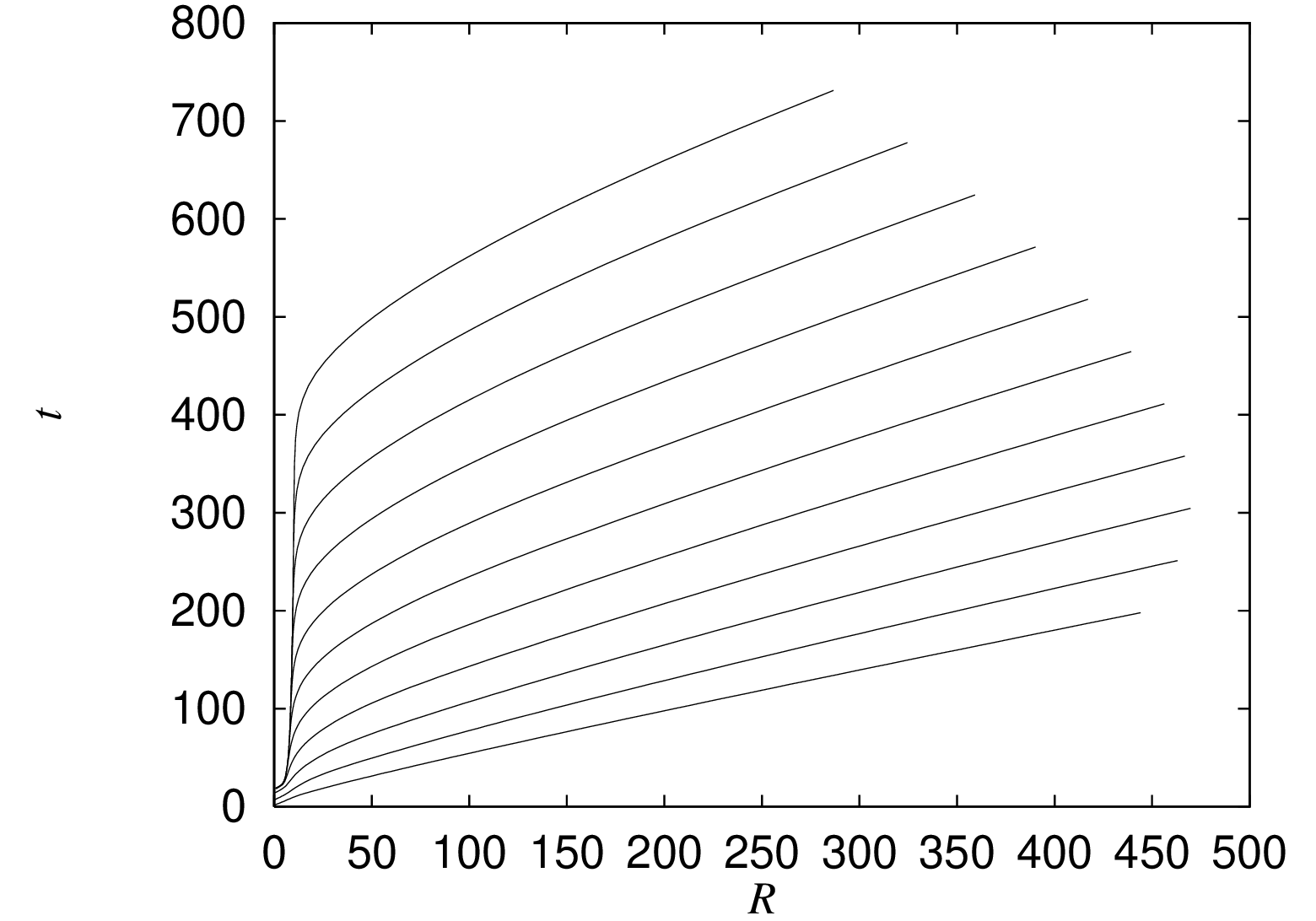}}
	\caption{Trajectories of outgoing light rays in the
        background geometry.}
	\label{fg:null}
\end{figure}

\begin{figure}[htbp]
	\centerline{\epsfxsize 10cm  \epsfysize 10cm \epsfbox{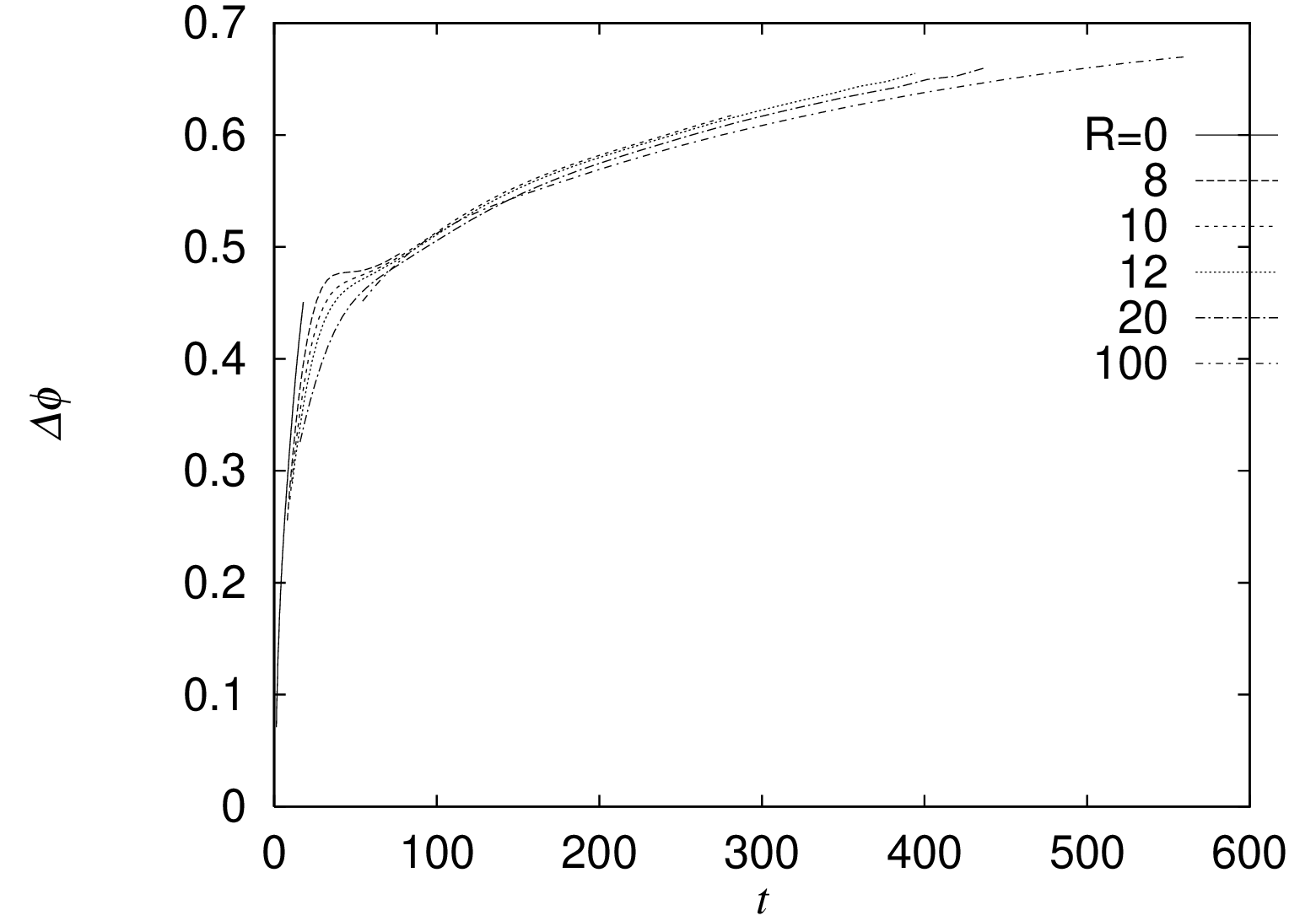}}
	(a)
	\centerline{\epsfxsize 10cm   \epsfysize 10cm  \epsfbox{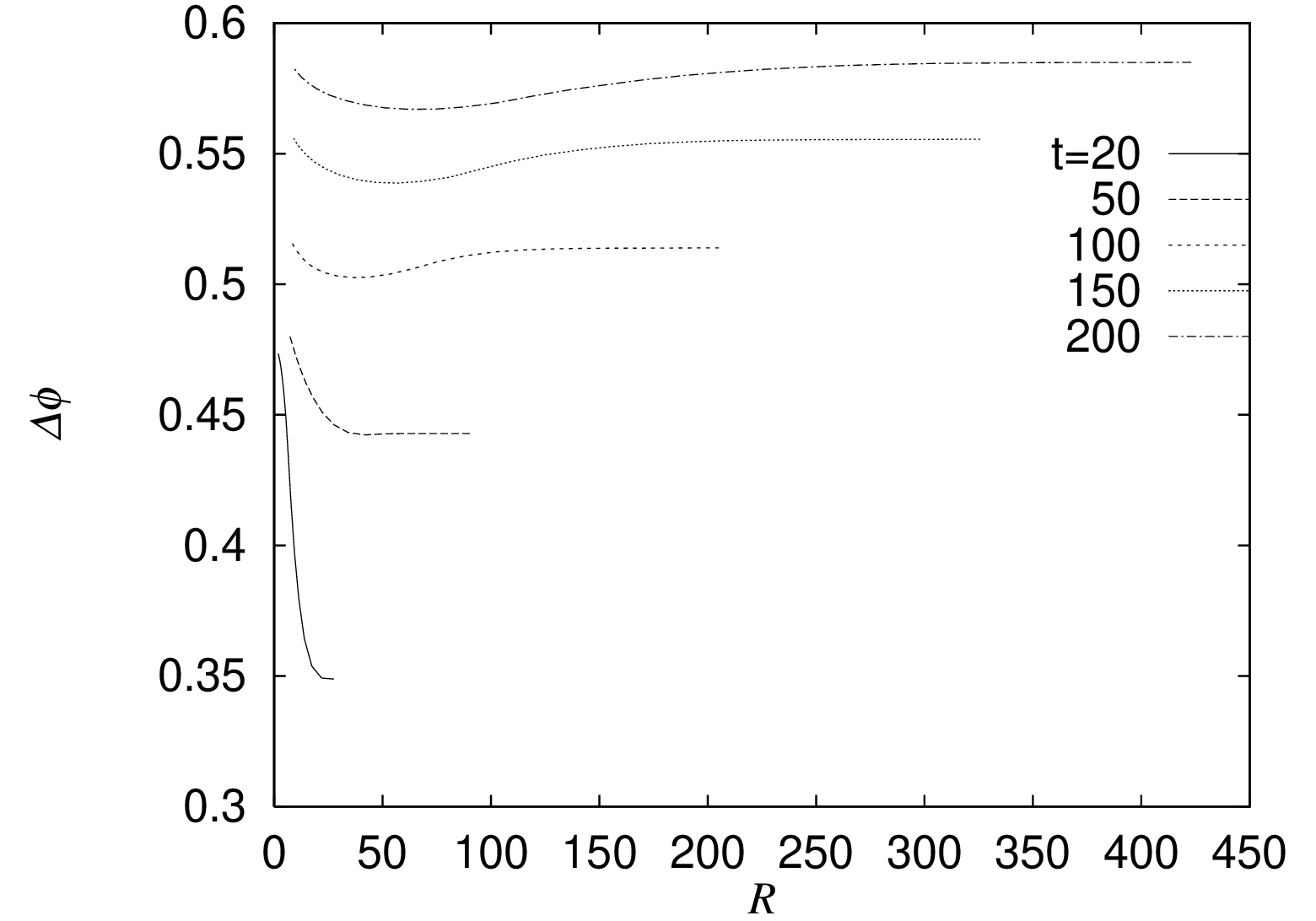}}	
	(b)
	\caption{$\Delta \phi\equiv (\phi-\phi_{0})/\phi_{0}$ 
        (a) along the world lines
	$R=\mbox{const}$ and on the hypersurfaces $t=\mbox{const}$.}
	\label{fg:phi}
\end{figure}

The results are shown in Figs.~\ref{fg:initial}-\ref{fg:phi}.
In Figs.~\ref{fg:initial} (a) and (b), 
a set of initial data for the background geometry
is shown.
The initial density perturbation $\delta$ is defined as
\begin{equation}
\delta(t,r)\equiv \frac{\rho(t,r)-\rho(t,\infty)}{\rho(t,\infty)}.
\end{equation}
In Fig.~\ref{fg:null}, the trajectories of 
outgoing light rays in this 
background geometry are shown.
It is shown that the event horizon is formed 
at $R\simeq 10$.

The results of integrating Eq.~(\ref{eq:BE}) are shown in
Figs.~\ref{fg:phi}(a) and (b), where
$\Delta\phi$ is defined as
\begin{equation}
\Delta\phi\equiv\frac{\phi(t,r)-\phi_{0}}{\phi_{0}}.
\end{equation}
It is found that the value of the scalar field around the black hole
is slightly smaller than the 
asymptotic value because of the underdensity of the surrounding region.
Nevertheless, the scalar field as a whole
is almost spatially homogeneous
at each moment, in spite of the existence of the black hole.
The evolution of the scalar field near the event horizon 
is well described by the cosmological homogeneous solution.

\section{Summary}
We have calculated the evolution of 
the Brans-Dicke scalar field in the 
presence of a primordial black hole
formed in a flat FRW universe.
We have found that 
the value of the scalar field 
at the event horizon almost maintains
the cosmological value at each moment.
This suggests that
primordial black holes ``forget''
the value of the gravitational constant
at their formation epoch.
However, it should be stressed that this result has only been
demonstrated for a dust universe in which the scalar
field does not appreciably affect the background curvature.
It remains to be seen whether the same conclusion applies when these
assumptions are dropped.

We are grateful to T.~Nakamura for helpful discussions
and useful comments.


\begin{thebibliography}{99}
\bibitem{hawking1975}
  S.W.~Hawking,
  Commun. Math. Phys.
  {\bf 43}, 199 (1975).
\bibitem{carr1975}
  B.J.~Carr,
  Astrophys. J.
  {\bf 201}, 1 (1975).
\bibitem{barrow1992}
  J.D.~Barrow,
  Phys. Rev. D
  {\bf 46}, 3227 (1992).
\bibitem{bc1996}
  J.D.~Barrow and B.J.~Carr,
  Phys. Rev. D
  {\bf 54}, 3920 (1996). 
\bibitem{cg2000}
  B.J.~Carr and C.A.~Goymer, 
  Prog. Theor. Phys.
  {\bf 136}, 321 (1999); 
   C.A.~Goymer and B.J.Carr,
   Phys. Rev. D., in press.
\bibitem{jacobson1999}
  T.~Jacobson,
  Phys. Rev. Lett.
  {\bf 83}, 2699 (1999).
\bibitem{will1999}
  T.M.~Eubanks et al., 
  Bull. Am. Phys. Soc., Abstract \#K 11.05 (1997); 
  C.M.~Will, 
  gr-qc/9811036. 
\bibitem{hcnn1997}
  T.~Harada, T.~Chiba, K.~Nakao and T.~Nakamura,
  Phys. Rev. D
  {\bf 55}, 2024 (1997).
\bibitem{inh1998}	
  H.~Iguchi, K.~Nakao and T.~Harada, 
  Phys. Rev. D
  {\bf 57}, 7262 (1998);
  H.~Iguchi, T.~Harada and K.~Nakao, 
  Prog. Theor. Phys. {\bf 101}, 1235 (1999);
  Prog. Theor. Phys. {\bf 103}, 53 (2000). 	
\bibitem{ch1974}
  B.J.~Carr and S.W.~Hawking,
  MNRAS
  {\bf 168}, 399 (1974).
\end{thebibliography}
\end{document}